\documentclass[twoside,leqno]{article}
\usepackage{ltexpprt}

\usepackage{color}
\usepackage{colortbl}
\usepackage{hhline}
\usepackage{amssymb}
\usepackage{graphicx}
\usepackage{url}
\usepackage{multirow}
\usepackage{morefloats}

\usepackage{siunitx}

\usepackage{mystyle_siam-compatible}

\begin{document}

\title{\Large Engineering Oracles for Time-Dependent Road Networks%
	\thanks{Partially supported by
		the EU FP7/2007-2013 under grant agreements no.~609026 (project MOVESMART) and no.~621133 (project HoPE), and
		by the DFG under grant agreements no.~WA654/16-2 and no.~WA654/23-1.
	\newline\vskip-5pt
	\indent
	$^\dag$ Computer Science \& Engineering Dept., University of Ioannina, 45110 Ioannina, GREECE.
	\protect{kontog@cs.uoi.gr}, \protect{gioulycs@gmail.com}.
	\newline
	\indent
	$^\ddag$ Computer Engineering \& Informatics Dept., University of Patras, 26500 Rio, GREECE.
	\protect{\{michalog,paraskevop,zaro\}@ceid.upatras.gr}.
	\newline
	\indent
	$^\diamond$ Computer Technology Institute and Press ``Diophantus'', 26504 Rio, GREECE.
	\newline
	\indent
	$^\between$ Karlsruhe Institute of Technology (KIT), Am Fasanengarten 5, 76131 Karlsruhe, GERMANY.
	%\newline
	\protect{dorothea.wagner@kit.edu}
	\newline\vskip-5pt
	\noindent
	Last change: November 26, 2015. Compiled: \today.
	}%THANKS
}%TITLE

\author%[Kontogiannis, Michalopoulos, Papastavrou, Paraskevopoulos, Wagner, Zaroliagis]
{%AUTHOR
				Spyros		Kontogiannis$^{\dag,\diamond}$
		\and	George		Michalopoulos$^{\ddag,\diamond}$
		\and 	Georgia	Papastavrou$^{\dag,\diamond}$
		\and 	Andreas	Paraskevopoulos$^{\ddag,\diamond}$
		\and	Dorothea	Wagner$^{\between}$
		\and	Christos	Zaroliagis$^{\ddag,\diamond}$
}%AUTHOR

\date{}

\maketitle

%\pagenumbering{arabic}
%\setcounter{page}{1}%Leave this line commented out.

\begin{abstract} 

\remove{%75-word abstract...
We implement and experimentally evaluate landmark-based oracles for min-cost paths in large-scale time-dependent road networks. We exploit parallelism and lossless compression, combined with a novel travel-time approximation technique, to severely reduce preprocessing space and time. We significantly improve the FLAT oracle, improving the previous query time by $30\%$ and doubling the Dijkstra-rank speedup. We also implement and experimentally evaluate a novel oracle (HORN), based on a landmark hierarchy, achieving even better performance wrt to FLAT.
}%REMOVE 75-word abstract...

%\remove{%LONG ABATRACT...

We implement and experimentally evaluate landmark-based oracles for min-cost paths in two different types of road networks with time-dependent arc-cost functions, based on distinct real-world historic traffic data: the road network for the metropolitan area of Berlin, and the national road network of Germany.

% FLAT ORACLE contribution...
Our first contribution is a significant improvement on the implementation of the $\alg{FLAT}$ oracle, which was proposed and experimentally tested in previous works. Regarding the implementation,
%% Parallelism...
we exploit \emph{parallelism} to reduce preprocessing time and real-time responsiveness to live-traffic reports.
%% Compression...
We also adopt a \emph{lossless compression scheme} that severely reduces preprocessing space and time requirements.
%% Landmark sets...
As for the experimentation, apart from employing the new data set of Germany, we also construct several \emph{refinements} and \emph{hybrids} of the most prominent landmark sets for the city of Berlin. A significant improvement to the speedup of $\alg{FLAT}$ is observed: 
%%%%%%%%%%%%%%%%%%%%%%%
% FLAT @ BERLIN 
%%%%%%%%%%%%%%%%%%%%%%%
For Berlin, the average query time can now be as small as $83\mathrm{\mu sec}$, achieving a speedup (against the time-dependent variant of Dijkstra's algorithm) of more than $1,119$ in absolute running times and more than $1,570$ in Dijkstra-ranks, with worst-case observed stretch less than $0.781\%$.
%%%%%%%%%%%%%%%%%%%%%%%
% FLAT @ GERMANY 
%%%%%%%%%%%%%%%%%%%%%%%
For Germany, our experimental findings are analogous: 
The average query-response time can be $1.269\mathrm{msec}$, achieving a speedup of more than $902$ in absolute running times, and  $1,531$ in Dijkstra-ranks, with worst-case stretch less than $1.534\%$.

% HORN ORACLE contribution...
Our second contribution is the implementation and experimental evaluation of a novel hierarchical oracle ($\alg{HORN}$). It is based on a \emph{hierarchy} of landmarks, with a few ``global'' landmarks at the top level possessing travel-time information for all possible destinations, and many more ``local'' landmarks at lower levels possessing travel-time information only for a small neighborhood of destinations around them. As it was previously proved, the advantage of $\alg{HORN}$ over $\alg{FLAT}$ is that it achieves query times sublinear, not just in the size of the network, but in the Dijkstra-rank of the query at hand, while requiring asymptotically similar preprocessing space and time. 
%%%%%%%%%%%%%%%%%%%%%%%
% HORN @ BERLIN 
%%%%%%%%%%%%%%%%%%%%%%%
Our experimentation of $\alg{HORN}$ in Berlin indeed demonstrates improvements in query times (more than $30.37\%$), Dijkstra-ranks (more than $39.66\%$), and also worst-case error (more than $35.89\%$), at the expense of a small blow-up in space.

% LIVE-TRAFFIC-UPDATES...
Finally, we implement and experimentally test a dynamic scheme to provide responsiveness to live-traffic reports of incidents with a small timelife (e.g., a temporary blockage of a road segment due to an accident). Our experiments also indicate that the traffic-related information can be updated in seconds.

%}%REMOVE LONG ABSTRACT...

\end{abstract}

%%%%%%%%%%%%%%%%%%%%%%%%%%%%%%%%%%%%%%%%%%%%%%%%%%%%%%%%%%%%%%%%
\section{Introduction}
\label{section:Introduction}

\term{Min-cost-path oracles} (a.k.a. distance oracles) are succinct data structures encoding min-cost path information among a carefully selected subset of pairs of vertices in a graph. The encoding is done in such a way that the oracle can efficiently answer min-cost path queries for arbitrary origin-destination pairs, exploiting the preprocessed data and/or local min-cost path searches. An oracle is exact (resp.~approximate) if the returned costs by the accompanying query algorithm are exact (resp.~approximate).
A bulk of important work (e.g., \cite{2005-Thorup-Zwick,2009-Sommer-Verbin-Yu,2010-Patrascu-Roditty,2011-Porat-Roditty,2012-Wulf-Nilsen-a,2012-Wulf-Nilsen-b,2013-Agarwal-Godfrey}) is devoted to constructing and analysing oracles for \emph{static} (i.e., \emph{time-independent}), mostly undirected networks in which the arc-costs are fixed scalars, providing trade-offs between the oracle's space and query time and, in case of approximate oracles, also of the approximation guarantee, typically tagged as the \emph{stretch} (factor). For an overview of oracles for networks with static arc-cost metrics, the reader is deferred to \cite{2014-Sommer-spq-survey}. %%% and references therein.

Considerable experimental work on routing in large-scale road networks has also appeared in recent years, with remarkable achievements that have been demonstrated on continental-size road-network instances. The goal is again to preprocess the arc-cost metric and then propose query algorithms (known as \term{speedup techniques} in this framework) for responding to min-cost path queries in time that is several orders of magnitude faster than the execution of Dijkstra's algorithm. An excellent overview of this line of research is provided in \cite{2014-Bast-Delling-Goldberg-Hannemann-Pajor-Sanders-Wagner-Werneck}. Once more, the bulk of the literature concerns \emph{static} metrics, with only a few exceptions \cite{bdsv-tdch-09, 2013-Batz-Geisberger-Sanders-Vetter,2011-Delling_TDSHARC,ndls-bastd-12}.

% WHY TIME DEPENDENCE?
Nevertheless, contemporary transportation networks are typically considered as directed graphs accompanied by an arc-cost metric (e.g., travel-time, energy-consumption) which demonstrates \emph{time-varying} behaviour.
For example, critical road segments in the center, or at an entry point of a metropolitan area, suffer traversal times which are heavily dependent on the actual time of traversal.
The rapid penetration of modern technology, such as the cellular technology, in our daily habits, has allowed travellers to act as moving sources of real-traffic information. Therefore, vehicle navigation and route planning vendors are nowadays in position to acquire in real time instantaneous live-traffic reports (e.g., road blockages due to car accidents), as well as periodic speed-probes that allow the maintenance of historic traffic data about this time-varying behaviour of each and every road segment in a road network.
This latter kind of historic traffic data can be succinctly represented as \emph{arc-traversal-time} functions, in particular the interpolants of the sampled arc-traversal-time values, for all the arcs in the network infrastructure.
One grand challenge is then to assess the earliest-arrival time for a given origin-destination pair $(o,d)\in V\times V$ and departure time $t_o\geq 0$ from the origin. If the metric of arc-traversal-times possesses the so-called \emph{FIFO} property, according to which a delayed departure  from $o$ cannot possibly lead to an earlier arrival at $d$, then all the classical shortest-path techniques (such as Dijkstra's and Bellman-Ford algorithms) have their time-dependent variants \cite{1969-Dreyfus,1990-Orda-Rom}. Of course, these are not the choices one should consider as query algorithms, in case of a route planning service that has to reply in real-time to several dozens, or even hundreds, of queries within a large-scale %%% road
network.

%%%%%%%%%%%%%%%%%%%%%%%%%%%%%%%%%%%%%%%%%%%%%%%%%%%%%%%%%%%%%%%%
\subsection{Related Work.}
\label{section:Related-Work}

Providing min-cost paths in graphs with time-dependent arc-costs is a hard problem. The FIFO property %%% indeed 
characterizes the complexity of the problem~\cite{1990-Orda-Rom}. For FIFO-abiding networks, the first time-dependent variant of Dijkstra's algorithm ($\alg{TDD}$) was proposed in~\cite{1969-Dreyfus}. %%% Actually, the equivalent model of non-FIFO networks with arbitrary waiting-at-nodes policy was considered in that work.

%%%% SPEEDUPS RELATED OVERVIEW %%%%%
A few time-dependent variants of well-known speedup techniques for road networks have appeared in the literature, with remarkable \emph{empirical} performance in some cases. All of them were experimentally evaluated on benchmark instances of the time-dependent European and/or the real data set for the German road networks, kindly provided by PTV AG for scientific use. In particular,
efficient approaches for responding to earliest-arrival-time queries are provided in \cite{2011-Delling_TDSHARC,2012-Delling-Nannicini}, by combining vertex contractions with goal-directed techniques.
The currently best speedup technique for both \emph{earliest-arrival-time} and \emph{travel-time function} (a.k.a. \emph{profile search}) computations in time-dependent road networks was given in \cite{2013-Batz-Geisberger-Sanders-Vetter}.
Their average response times to arbitrary queries for the German road network, of $4.7M$ vertices and $10.8M$ arcs, is less than $1.5\mathrm{msec}$ and preprocessing space requirements of less than $1GB$. A point-to-point travel-time function can be constructed in less than $40\mathrm{msec}$, when the departure times interval is a single day. For point-to-point approximate travel-time functions, with experimentally observed stretch at most $1\%$, the construction time is less than $3.2\mathrm{msec}$. Their approach is based on the \emph{time-dependent Contraction Hierarchies} \cite{bdsv-tdch-09}, along with several heuristic improvements on the preprocessing step and on the query method.

To our knowledge, oracles for time-dependent networks achieving \emph{provable} approximation guarantees, small preprocessing-space complexity and sublinear query time complexity, have been only recently proposed and theoretically analysed \cite{2014-Kontogiannis-Zaroliagis,2015-Kontogiannis-Wagner-Zaroliagis}, and experimentally evaluated \cite{2015-Kontogiannis-Michalopoulos-Papastavrou-Paraskevopoulos-Wagner-Zaroliagis}.

%%%%%%%%%%%%%%%%%%%%%%%%%%%%%%%%%%%%%%%%%%%%%%%%%%%%%%%%%%%%%%%%
\subsection{Our Contribution.}
\label{section:Contribution}

We engineer and experimentally evaluate oracles for time-dependent road networks on two real instances corresponding to qualitatively different cases. 
The first one corresponds to a typical weekday of the metropolitan road network of Berlin (about $4.7K$ vertices and $1.13M$ arcs) and was kindly provided to us by TomTom \cite{tomtom} for scientific use. The second instance corresponds to a typical weekday of the German road network (about $4.6M$ vertices and $11.18M$ arcs) and was kindly provided to us by \cite{ptv}, also for scientific use. More details on the identities of the two instances can be found in Table~\ref{table:arc-travel-time-statistics}.

In this work, we make two main contributions. First, we significantly extended and improved the implementation and experimental evaluation of the $\alg{FLAT}$ oracle proposed in \cite{2014-Kontogiannis-Zaroliagis} and originally implemented and experimentally assessed in \cite{2015-Kontogiannis-Michalopoulos-Papastavrou-Paraskevopoulos-Wagner-Zaroliagis}. Regarding the implementation, we exploit \emph{parallelism} to reduce preprocessing time and real-time responsiveness to live-traffic reports. We also adopt a \emph{lossless compression scheme} that severely reduces preprocessing space and time requirements.
As for the experimentation, apart from employing the new data set of Germany, we also construct several \emph{refinements} and \emph{hybrids} of the most prominent landmark sets (\textsc{random} and \textsc{kahip}) for the city of Berlin. A significant improvement to the speedup of $\alg{FLAT}$ is observed: the average query time can now be as small as $81\mathrm{\mu sec}$, if we exploit the query algorithm $\alg{FCA}$ along with the \textsc{sparse-random} landmark set, corresponding to a speedup against $\alg{TDD}$ that is larger than $1,146$ w.r.t. query-response times and more than $1,227$ in Dijkstra-ranks, with worst-case observed stretch less than $0.771\%$, for the Berlin instance. If we use the \textsc{sparse-kahip} landmark set, the absolute query time is slightly worse ($83\mathrm{\mu sec}$) but the speedup w.r.t. the Dijkstra-ranks is now more than $1,570$, with worst-case error less than $0.781\%$.

Our experimental findings for Germany are analogous: the average response-time to arbitrary queries can be at most $1.269\mathrm{msec}$, achieving a speedup of more than $902$ in absolute running times and $1,531$ in Dijkstra-ranks, with worst-case stretch at most $1.534\%$.

Our second contribution is the implementation and experimental evaluation of the novel $\alg{HORN}$ oracle \cite{2015-Kontogiannis-Wagner-Zaroliagis},
which is based on a \emph{hierarchy} of landmarks, from a few ``global'' landmarks possessing knowledge of the entire network towards (many more) ``local'' landmarks whose knowledge of the network is restricted to small neighborhoods around them. As was proved in \cite{2015-Kontogiannis-Wagner-Zaroliagis}, the advantage of $\alg{HORN}$ over $\alg{FLAT}$ is that it achieves query times \emph{sublinear}, not just in the size of the network, but in the actual Dijkstra-rank of the query at hand, be it long-range, mid-range, or short-range, while requiring similar preprocessing space and time. Our experiments on the Berlin instance indeed confirm the improved stretch factors, but also better speedups due to sophisticated early-stopping criteria, compared to the experimentation on $\alg{FLAT}$ for the same subsets of ``global'' landmarks. In particular, our experimentation of $\alg{HORN}$ in Berlin indeed demonstrates improvements in query times (more than $30.37\%$), Dijkstra-ranks (more than $39.66\%$), and also worst-case error (more than $35.89\%$), at the expense of a small blow-up in space.

An additional contribution concerns the implementation and experimental evaluation of a dynamic scheme to provide responsiveness to live-traffic reports of incidents with a small timelife (e.g., a temporary blockage of a road segment due to an accident). Our experiments also indicated that traffic-related information can be updated in seconds.

%%%%%%%%%%%%%%%%%%%%%%%%%%%%%%%%%%%%%%%%%%%%%%%%%%%%%%%%%%%%%%%%
\section{Preliminaries}
\label{section:Preliminaries}

%%%%%%%%%%%%%%%%%%%%%%%%%%%%%%%%%%%%%%%%%%%%%%%%%%%%%%%%%%%%%%%%
\subsection{Notation.}
\label{section:notation}

We consider directed graphs $G=(V,A)$ with $|V| = n$ vertices and $|A|=m=\Order{n}$ arcs,
where each arc $a\in A$ is accompanied with a continuous, periodic, piecewise linear (pwl) \term{arc-travel-time} (or \term{arc-delay}) function defined as follows: $\forall k\in \naturals, \forall t\in [0,T),~ D[a](kT+t) = d[a](t)$, where $d[a]: [0,T) \rightarrow [1,M_a]$ such that $\lim_{t\uparrow T}d[a](t) = d[a](0)$, for some fixed integer $M_a$ denoting the maximum possible travel time ever observed at arc $a$. Notice that the minimum arc travel time value in the entire network is also normalized to $1$. Each arc-travel-time function $D[a]$ can be represented succinctly as a list of $K_{a}$ breakpoints defining $d[a]$.
Let
	$K=\sum_{a\in A} K_{a}$
be the number of breakpoints to represent all of them,
	$K_{\max} = \max_{a\in A} K_{a}$, and
	$K^*$ be the number of \emph{concavity-spoiling} breakpoints, i.e., those in which the arc-travel-time slopes increase.
Clearly, $K^* \leq K$, and $K^* = 0$ for \emph{concave} pwl arc-travel-time functions.

The \term{arc-arrival-time} functions are defined as $Arr[a](t) = t + D[a](t)$, $\forall t\in [0,\infty)$.
% FIFO Property...
	A typical assumption is that each arc-arrival-time functions are strictly increasing, in order to satisfy the strict FIFO property.
The \term{path-arrival-time} function of a path $p = \langle a_1,\ldots,a_k \rangle$ in $G$ (represented as a sequence of arcs) is defined as the composition of the arc-arrival-time functions for the constituent arcs of $p$:
\(
	Arr[p](t) = Arr[a_k](Arr[a_{k-1}](\cdots(Arr[a_1](t))\cdots))\,.
\)
The \emph{path-travel-time} function is $D[p](t) = Arr[p](t) - t$. Also, between any origin-destination $(o,d)\in V\times V$, $\mathcal{P}_{o,d}$ is the set of all $od-$paths in $G$, and the \term{earliest-arrival-time} / \term{shortest-travel-time} functions are defined as follows: $\forall t_o\geq 0$,
\(
	Arr[o,d](t_o)
	= \min_{p\in \mathcal{P}_{o,d}}\left\{ Arr[p](t_o)\right\}
\)
and
\(
	D[o,d](t_o)
	= \min_{p\in \mathcal{P}_{o,d}}\left\{ D[p](t_o)\right\} = Arr[o,d](t_o) - t_o\,.
\)
The \term{Dijkstra-rank} $\Gamma[o,d](t_o)$ is the number of settled vertices up to $d$, when executing the time-dependent variant of $\alg{Dijkstra}$ (we call it $\alg{TDD}$) from $(o,t_o)$. $SP[o,d](t_o) = \{ p\in P_{o,d} : Arr[p](t_o) = Arr[o,d](t_o) \}$ is the set of minimum-travel-time paths for the triple $(o,d,t_o)$. $ASP[o,d](t_o)$ is the set of $od$-paths which are $(1+\eps)$-approximations of minimum-travel-time $od$-paths.

%%%%%%%%%%%%%%%%%%%%%%%%%% FREE-FLOW & FULL-CONGESTION METRICS %%%%%%%%%%%%%%%%%%%%%%%%%%
For any $a=uv\in A$ and subinterval $[t_s,t_f)\subseteq [0,T)$, we define upper-bounding and lower-bounding (static) travel-time metrics:
%
%\begin{itemize}
	%
	%\item
	the \term{free-flow} travel-time $\underline{D}[uv](t_s,t_f) := \min_{t_u\in [t_s,t_f)} \{ D[uv](t_u) \}$  and
	%
	%\item
	the \term{maximally-congested} travel-time $\overline{D}[uv](t_s,t_f) := \max_{t_u\in [t_s,t_f)} \{ D[uv](t_u) \}$.
%
%\end{itemize}
%
If $[t_s,t_f) = [0,T)$, we refer to the static \term{free-flow} and \term{full-congestion} metrics $\underline{D} , \overline{D} : A \rightarrow [1,M]$, respectively. With respect to any arc-cost metric $D$, $diam(G,D)$ is the diameter of the graph, i.e., the largest possible distance in the graph. For example, $diam(G,\underline{D})$ is the free-flow diameter of $G$.

%%%%%%%%%%%%%%%%%%%%%%%% DEFINITIONS OF DIJKSTRA BALLS %%%%%%%%%%%%%%%%%%%%%%%%
For any vertex $v\in V$, departure-time $t_v\in\nonnegativereals$ and integer $F\in[n]$, $B[v;F](t_v)$ ($B[v;R](t_v)$) is a ball of size $F\in \naturals$ (of radius $R>0$) growing by $\alg{TDD}$ from $(v,t_v)$ in the time-dependent metric.
Analogously, $\underline{B}[v;F]$ ($\underline{B}[v;R]$), and $\overline{B}[v;F]$ ($\overline{B}[v;R]$) are the size-$F$ (radius-$R$) balls from $v$ under the (static) free-flow and fully-congested travel-time metrics.

%%%%%%%%%%%%%%%%%%%%%%%% UPPER-APPROXIMATING DISTANCE FUNCTIONS %%%%%%%%%%%%%%%%%%%%%%%%
A \term{$(1+\eps)$-upper-approximation} $\overline{\De}[o,d]$ and a \term{$(1+\eps)$-lower-approximation} $\underline{\De}[o,d]$ of $D[o,d]$, are continuous, pwl, periodic functions, with a (hopefully small) number of breakpoints in $[0,T)$, such that:
%\begin{equation}
%	\label{eq:upper-approximating-minimum-travel-time}
	$\forall t_o\geq 0,~
	 D[o,d](t_o) ~/~ (1+\eps)
	\leq \underline{\De}[o,d](t_o)
	\leq D[o,d](t_o)
	\leq \overline{\De}[o,d](t_o)
	\leq (1+\eps)\cdot D[o,d](t_o)\,.$
%\end{equation}

For convenience, the notation used throughout the paper is summarized in Table~\ref{table:notation}.

%%%%%%%%%%%%%%%%%%%%%%%%%%%%%%%%%%%%%%%%%%%%%%%%%%%%%%%%%%%%%%%%
\subsection{Assumptions on Travel-Time Metric.}
\label{section:assumptions}

We adopt two assumptions from \cite{2014-Kontogiannis-Zaroliagis} and an additional
one from \cite{2015-Kontogiannis-Wagner-Zaroliagis} on the kind of shortest-travel-time functions that may appear in the time-dependent network instance at hand. All of them are quite natural and justified in time-depenent road networks. 
Technically, these assumptions allow the \emph{smooth transition} from static metrics on undirected graphs towards time-dependent metrics on directed graphs.
% NARRATION OF ASSUMPTIONS...
The first assumption asserts that all the minimum-travel-time slopes are bounded in a given interval $[-\La_{\min},\La_{\max}]$.

%%%%%%%%%%%%%%%%%%%%%%%%%%%%%%%%%%%%%%%%%%%%%%%%%%%%%%%%%%%%%%%%
\begin{assumption}
\label{assumption:Bounded-Travel-Time-Slopes}
	$\exists\La_{\min}\in[0,1)$, $\exists\La_{\max}\geq 0$ s.t. the following holds:
\(
	\forall (o,d)\in V\times V,~ \forall 0\leq t_1 < t_2,~
	(D[o,d](t_1) - D[o,d](t_2)) ~/~ (t_1 - t_2) \in[-\La_{\min}, \La_{\max}]\,.
\)
\end{assumption}
%%%%%%%%%%%%%%%%%%%%%%%%%%%%%%%%%%%%%%%%%%%%%%%%%%%%%%%%%%%%%%%%

The lower-bound of $-1$ in the minimum-travel-time function slopes is indeed a direct consequence of the FIFO property, which is typically assumed to hold in several time-dependent networks, such as road networks.
$\La_{\max}$ represents the maximum possible rate of change of minimum-travel-times in the network, which only makes sense to be bounded (in particular, independent of the network size) in realistic instances such as the ones representing urban-traffic time-dependent road networks.

The next assumption states that the ratio of minimum-travel-times in opposite directions between two vertices, for any departure-time but not necessarily via the same path, is bounded by a constant.

%%%%%%%%%%%%%%%%%%%%%%%%%%%%%%%%%%%%%%%%%%%%%%%%%%%%%%%%%%%%%%%%
\begin{assumption}
\label{assumption:Bounded-Opposite-Trips}
$\exists\zeta \geq 1$,
\(
	\forall (o,d)\in V\times V,~ \forall t\in[0,T),~
	D[o,d](t) \leq \zeta\cdot D[d,o](t)\,.
\)
\end{assumption}
%%%%%%%%%%%%%%%%%%%%%%%%%%%%%%%%%%%%%%%%%%%%%%%%%%%%%%%%%%%%%%%%
%

The last assumption states that the free-flow ball from a vertex blows-up by at most a polylogarithmic factor, if it is expanded up to the full-congestion radius within it, under the free-flow metric.

%%%%%%%%%%%%%%%%%%%%%%%%%%%%%%%%%%%%%%%%%%%%%%%%%%%%%%%%%%%%%%%%
\begin{assumption}
\label{assumption:growth-of-free-flow-ball-sizes}
For any vertex $\ell\in V$, and a positive integer $F$, consider the (static) Dijkstra ball $\underline{B}[\ell;F]$ around $\ell$ under the free-flow metric.
Let
	$\underline{R}[\ell] = \max\{ \underline{D}[\ell,v]: v\in \underline{B}[\ell;F] \}$
and
	$\overline{R}[\ell] = \max\{ \overline{D}[\ell,v]: v\in \underline{B}[\ell;F] \}$
be the largest free-flow and full-congestion travel-times from $\ell$ to any other vertex in $\underline{B}[\ell;F]$.
Finally, let
	$\underline{B}'[\ell;F] = \{ v\in V: \underline{D}[\ell,v](0,T) \leq \overline{R}[\ell] \}$
be the free-flow ball around $\ell$ with the (larger) radius $\overline{R}[\ell]$.
Then it holds that $|\underline{B}'[\ell;F]| \in \Order{F\cdot\polylog(F)}$.
\end{assumption}
%%%%%%%%%%%%%%%%%%%%%%%%%%%%%%%%%%%%%%%%%%%%%%%%%%%%%%%%%%%%%%%%

We conducted extended experimental analysis for two distinct road networks with time-dependent travel-times, the metropolitan area of Berlin, and the national road network of Germany. The former was kindly provided to us by TomTom~\cite{tomtom} and the latter by PTV AG~\cite{ptv}, for scientific purposes. More details on the identities of our instances can be found in Table~\ref{table:arc-travel-time-statistics}.
We verified the validity of these assumptions (cf.~Tables~\ref{table:arc-travel-time-statistics}, \ref{table:Path-Travel-Time-Statistics}, and \ref{table:Free-Flow-Rank-Blowup}). The maximum observed values for $\zeta$, $\La_{\max}$ and the blow-up factor in free-flow ball size are $1.189$, $0.216$ and $8.3$, respectively. More details are provided in Section~\ref{section:identity-of-instances}.

%%%%%%%%%%%%%%%%%%%%%%%%%%%%%%%%%%%%%%%%%%%%%%%%%%%%%%%%%%%%%%%%
\section{Time-Dependent Oracles}
\label{section:Time-Dependent-Oracles}

%%%%%%%%%%%%%%%%%%%%%%%%%%%%%%%%%%%%%%%%%%%%%%%%%%%%%%%%%%%%%%%%
\subsection{The FLAT Oracle.}

The main rationale of the $\alg{FLAT}$ oracle %%% , presented and analysed in
\cite{2015-Kontogiannis-Wagner-Zaroliagis}
is the following: we somehow select a small subset of landmark vertices $L$.
Then, we compute \term{travel-time summaries}, i.e., $(1+\eps)-$approximations of minimum-travel-time \emph{functions}, from landmarks towards all reachable vertices. Finally, we use properly designed query algorithms that exploit these
travel-time summaries in order to provide approximately minimum-travel-time \emph{values} for arbitrary queries $(o,d,t_o)\in V\times V\times [0,T)$.

%%%%%%%%%%%%%%%%%%%%%%%%%%%%%%%%%%%%%%%%%%%%%%%%%%%%%%%%%%%%%%%%
\subsubsection{Selection of Landmark Sets.}
$\alg{FLAT}$ was preliminarily implemented and evaluated in
\cite{2015-Kontogiannis-Michalopoulos-Papastavrou-Paraskevopoulos-Wagner-Zaroliagis}
for six landmark sets of the Berlin instance. We considered three selection methods
(\textsc{random}, \textsc{metis}~\cite{2014-METIS} and \textsc{kahip}~\cite{2014-KAHIP}) and two sizes ($1,000$ and $2,000$ landmarks).
The experimental evaluation in \cite{2015-Kontogiannis-Michalopoulos-Papastavrou-Paraskevopoulos-Wagner-Zaroliagis} indicated that the \textsc{random} (R) and \textsc{kahip} (K) landmark sets dominated the rest, the former w.r.t. the query times and the latter w.r.t. observed relative errors. In this work we choose to focus on these two prominent landmark sets.

For the Berlin instance, we create landmark sets, each containing $2,000$ landmarks. The space requirements were originally quite demanding. After the lossless compression techniques that we adopt here (cf. Section~\ref{section:Compressing-Preprocessing-Space}) we managed to decrease the space by 2/3.
% REFINEMENTS OF \textsc{random} LANDMARKS...
We consider also some refinements of the R- and K-landmark sets:

\begin{itemize} 

\item \textsc{important-random} (IR) landmarks: We take as input the R-landmark set and from each landmark we grow a small ball in the free-flow metric. Within this ball we move the landmark from the ball center to a vertex adjacent to the most significant road segment in the ball.

\item \textsc{sparse-random} (SR) landmarks: We choose the landmarks again randomly but sequentially: Each time a landmark is selected, a small ball is subtracted from the network prior to the selection of the next landmark, so as to assure that the landmarks are well separated.

\item \textsc{sparse-kahip} (SK) landmarks: We initially choose a superset of candidate-landmarks from a fine-grained KaHIP partition. The final landmark set is chosen from these candidates, once more randomly and sequentially: Each time a landmark is selected, a small ball of candidate-landmarks around it is also subtracted prior to the selection of the next landmark, so as to assure that the landmarks are well separated.

\item \textsc{hybrid} (H) landmarks: This landmark selection method combines the rationales of \textsc{random} and \textsc{kahip}. We start with a \textsc{kahip} partition that provides roughly $1,000$ boundary vertices as landmarks. Another collection of roughly $1,000$ additional landmarks is selected, in a balanced way among cells and uniformly at random within each cell of the \textsc{kahip} partition.

\end{itemize}

%HIERARCHICAL LANDMARKS...
Finally, for the sake of the $\alg{HORN}$ oracle, we also construct hierarchical landmark sets, \textsc{hierarchical}-\textsc{random} (HR) and \textsc{hierarchical}-\textsc{sparse}-\textsc{random} (HSR) of $10,256$ and $20,513$ landmarks.

%GERMANY LANDMARKS
For the instance of Germany we only consider the R-, K-, SR- and SK-landmark sets, each of roughly $2,000$ landmarks. Unfortunately we could not construct an IR-landmark set to this instance, since there is no such information concerning the classification of road segments.

%%%%%%%%%%%%%%%%%%%%%%%%%%%%%%%%%%%%%%%%%%%%%%%%%%%%%%%%%%%%%%%%
\subsubsection{Preprocessing Phase.}
After $L$ is determined, a preprocessing phase is performed in which all
$(1+\eps)-$upper-approximating travel-time functions (travel-time summaries) from
landmarks $\ell\in L$ towards destinations $\forall v\in V$ are computed and stored,
based solely on the $\alg{TRAP}$ method \cite{2015-Kontogiannis-Wagner-Zaroliagis}.
$\alg{TRAP}$ splits the entire period $[0,T)$ into small, consecutive subintervals,
each of length $\tau>0$, and provides a crude approximation of the unknown
shortest-travel-time function in each interval, based solely on
Assumption~\ref{assumption:Bounded-Travel-Time-Slopes}. %%%% concerning the boundedness of the shortest travel-time slopes in the instance.
After sampling the travel-time values of each destination $v\in V$, for a given origin $u\in V$,
we consider each pair of consecutive sampling times $t_s < t_f$ and the semilines with
slopes $\La_{\max}$ from $t_s$ and $-\La_{\min}$ from $t_f$. The considered upper-approximating
function $\upperD[u,v]$ within $[t_s,t_f)$ is then (a refinement of) the lower-envelope of these two lines.

%%%%%%%%%%%%%%%%%%%%%%%%%%%%%%%%%%%%%%%%%%%%%%%%%%%%%%%%%%%%%%%%
\subsubsection{Query Algorithms.}

We consider the three query algorithms $\alg{FCA}$, $\alg{RQA}$, and $\alg{FCA}^+(N)$. The first two were introduced in \cite{2014-Kontogiannis-Zaroliagis}, while the third one was introduced in \cite{2015-Kontogiannis-Michalopoulos-Papastavrou-Paraskevopoulos-Wagner-Zaroliagis}. All algorithms can be fine-tuned to run in $o(n)$ time. A preliminary implementation and experimental evaluation was provided in \cite{2015-Kontogiannis-Michalopoulos-Papastavrou-Paraskevopoulos-Wagner-Zaroliagis}.

$\alg{FCA}$ grows a ball $B_o \equiv B[o](t_o) = \left\{x\in V : D[o,x](t_o) \leq D[o,\ell_o](t_o)\right\}$ from $(o,t_o)$, by running $\alg{TDD}$ until either $d$ or the closest landmark $\ell_o \in \arg\min_{\ell\in L}\{ D[o,\ell](t_o)\}$ is settled. It then returns either the exact travel-time value, or an ($1+\eps+\psi$)-approximate travel-time value via $\ell_o$, where $\psi$ is a constant depending on $\eps, \zeta$, and $\La_{\max}$, but not on $n$.

$\alg{FCA}^+(N)$ is a variant of $\alg{FCA}$ that keeps growing a $\alg{TDD}$ ball from $(o,t_o)$ until either $d$, or a given number $N$ of landmarks is settled, and then returns the smallest via-landmark approximate travel-time value (among all $N$ settled landmarks). The approximation guarantee is the same as that of $\alg{FCA}$, but its practical performance is impressive (in most cases even better than $\alg{RQA}$).

$\alg{RQA}$ is a PTAS, providing an approximation guarantee of 
$1+\s 
= 1 + \eps\cdot\left[(1+\eps/\psi)^{r+1}\right] / \left[(1+\eps/\psi)^{r+1} - 1 \right]$, 
by exploiting carefully a number $r\in \naturals$ (called the \term{recursion budget}) of recursive accesses to the preprocessed information, each of which produces (via calls to $\alg{FCA}$) additional candidate $od-$paths $sol_i$.
$\alg{RQA}$ works as follows. As long as the destination vertex within the explored area around the origin has not yet been discovered, and there is still some remaining recursion budget, it ``guesses'' (by exhaustively searching for it) the next vertex $w_k$ of the boundary set of touched vertices (i.e., still in the priority queue) along the unknown shortest $od-$path. Then it grows an outgrowing $\alg{TDD}$ ball from the new center $(w_k, t_k = t_o+D[o,w_k](t_o))$, until it reaches the closest landmark $\ell_k$ to it, at travel-time $R_k = D[w_k,\ell_k](t_k)$. This new landmark offers an alternative $od-$path $sol_k$ by a new application of $\alg{FCA}$.

% DIFFERENT EXPERIMENTATION OF FCA+...
In \cite{2015-Kontogiannis-Michalopoulos-Papastavrou-Paraskevopoulos-Wagner-Zaroliagis} we executed $\alg{FCA}^+(N)$ \emph{after} the execution of $\alg{RQA}$, so as to set the value of $N$ to the actual number of landmarks that were discovered by $\alg{RQA}$, in each query. To avoid this privileged treatment of $\alg{FCA}^+(N)$, we choose here to test it for several  but fixed values of $N$, regardless of $\alg{RQA}$'s behavior.

%%%%%%%%%%%%%%%%%%%%%%%%%%%%%%%%%%%%%%%%%%%%%%%%%%%%%%%%%%%%%%%%
\subsection{The HORN Oracle.}

The novelty of the $\alg{HORN}$ oracle \cite{2015-Kontogiannis-Wagner-Zaroliagis}
is to create a hierarchy of landmark sets, whose range of ``preprocessed destinations'' gradually ranges from a few ``nearby'' vertices up to all reachable vertices (in the last level), in order to serve each query $(o,d,t_o)$ only with relevant landmarks with respect to its own Dijkstra-rank $\Gamma[o,d](t_o)$. This way, we aim at achieving speedups similar to those of $\alg{FLAT}$ for long-range queries, to all possible ranges of queries, while increasing the space requirements only by a small factor. Our goal is to ``guess'' the order of the Dijkstra Rank $\Gamma[o,d](t_o)$ for $(o,d,t_o)$. The guessing is achieved in a way that is typical in online algorithms that have to deal with an unknown parameter: Starting from a small value (say, $\Order{\sqrt{n}}$), we keep growing a ball from $(o,t_o)$, increasing appropriately the value of the guess as the ball grows, until the very first time at which a successful completion of a proper variant of $\alg{RQA}$ is very likely to occur (exactly because we ``guessed right'' the actual Dijkstra-rank). The travel-time returned is that of the best possible $od$-path among all the successfully discovered approximate $od$-paths so far, via ``informed'' landmarks that possess travel-time summaries for $d$. The crux is in organizing the preprocessed information in such a way that it is indeed possible for the query algorithm to successfully complete its execution as soon as the ``guess'' asymptotically matches the value of $\Gamma[o,d](t_o)$.

% DESCRIPTION OF HQA...
The \term{Hierarchical Query Algorithm} ($\alg{HQA}$) for $(o,d,t_o)$ proceeds as follows: a single ball grows from $(o,t_o)$, until either $d$ is reached, or an \emph{Early Stopping criterion} (ESC) is fulfilled, or the \emph{Appropriate Level of Hierarchy} (ALH) of landmarks is reached (whichever occurs first).
If $d$ is settled by the ball from $(o,t_o)$, an exact solution is returned.
% ESC occurs ...
If ESC causes $\alg{HQA}$ to terminate, then the value $D[o,\ell_o](t_o) + \overline{\De}[\ell_o,d](t_o+D[o,\ell_o](t_o))$ is reported, because it is already a very good approximation.
% ALH holds ...
Otherwise, $\alg{HQA}$, due to ALH, considers being at the right level-$i$ of the hierarchy and continues executing the corresponding variant of $\alg{RQA}$, call it $\alg{RQA}_i$, which uses as its own landmark set $M_i = \union_{j=i}^{4} L_j$. Observe that $\alg{RQA}_i$ may now fail constructing approximate shortest paths via certain landmarks in $M_i$ that it settles, since they may not possess a travel-time summary for $d$. $\alg{HQA}$ terminates by returning the best $od$-path that has been discovered so far, via \emph{all} settled landmarks which are ``informed'' (i.e., they have $d$ in their coverage), either by the very first ball from $(o,t_o)$ or by $\alg{RQA}_i$.
$\alg{HQA}$ uses some parameters: $a$ is the degree of sublinearity in the query time, compared to the targeted Dijkstra-rank; $\beta$ is related to the approximation guarantee achieved upon exit due to ESC; $\gamma$ has to do with the number of levels that we create in the hierarchy; and $\xi$ is the amount of slackness that we introduce in the size of the area of coverage. We set these parameters here to the values $a = 1,~ \beta = 1,~ \gamma = 1.88,~ \xi = 0.1$. A more detailed explanation of $\alg{HQA}$, as well as of its parameters, is provided in \cite{2015-Kontogiannis-Wagner-Zaroliagis}.

%%%%%%%%%%%%%%%%%%%%%%%%%%%%%%%%%%%%%%%%%%%%%%%%%%%%%%%%%%%%%%%%
\section{Compressing Preprocessing Space}
\label{section:Compressing-Preprocessing-Space}

Due to the criticality of the preprocessing space, our goal is to achieve an efficient
storage of the constructed travel-time summaries, while keeping a sufficient precision. The key is that some specific features can be exploited in order to reduce the required space.

\subsection{Fixed Range.}
For a one-day time period, departure-times and arrival-times have a bounded value range. The same also holds for travel times which are at most one-day for any query within a country area such as Germany. Therefore, when the considered precision of the traffic data is within seconds, we handle time-values as integers in the range $[0 ~,~ 86,399]$, for milliseconds as integers in $[0~,~ 86,399,999]$, etc.
Any (real) time value within a single-day period, represented as a floating-point number $t_f$, can thus be converted to an integer $t_i$ with fewer bytes and a given unit of measure. For a unit measure (or scale factor) $s$, the resulting integer is $t_i = \lceil t_f / s \rceil$. In this manner, $t_i$ needs size $\lceil \log_2(t_f / s) / 8 \rceil$ bytes. The division $t_f / s$ has quotient $\pi$ and remainder $\upsilon$. Thus, $t_f = s \cdot \pi + \upsilon$ and $t_i = \lceil (s \cdot \pi + \upsilon) / s \rceil = \lceil \pi + \upsilon / s \rceil$, with $\upsilon < s$. Therefore, converting $t_f$ to $t_i$ results to an absolute error of at most $2s$. In the reverse process, for extracting the stored value, the conversion is $t_{f}^{'}= t_i \cdot s$.

\subsection{Bucketing.}
The number of breakpoints of the arc-travel-time functions is a major factor of space increase on the resulting minimum-travel-time functions. A way to deal with this is by merging consecutive breakpoints having absolute difference in travel-time values less than few seconds, in each arc travel time function. In this manner, to preserve the upper bound error, each resulted breakpoint gets the largest travel time among the breakpoints which take part in merge. Depending on the bucketing parameter $c$, we can decrease the number of breakpoints, sacrificing part of accuracy. In our experiments, the bucketing led to the highest reduction (about $86\%$) in space requirements.

\subsection{Piecewise Composition.}
Many shortest $od$-paths typicaly contain at least one arc with pwl travel-time function, making $D[o,d](t)$ also a pwl function. To avoid the space increase from storing breakpoints unrestrictedly, we analyse any such shortest $od$-path into two subpaths $o$-$p$-$d$. $p$ is selected so that the $pd$-subpath is the maximum subpath with no arc having pwl travel-time function. Such $pd$-subpaths exist because the number of constant arc-travel-time functions is much larger than the number of the pwl ones. Thus, for $D[o,d](t)$ we only store a ``predecessor'' pointer to $D[o,p](t)$ and the \emph{constant} travel-time offset $D[p,d]$ i.e., $D[o,d](t)=D[o,p](t) + D[p,d]$. In our experiments, this method lead to $40\%$ reduction of the space requirements.

\subsection{Delay Shifts.}
There are many shortest $od$-paths with travel-time $D[o,d](t)$ with delay variation. Thus we further reduced the required space as follows. The delay fluctuates around a constant value. By taking the minimum delay value, the leg-delays can be represented as small shifts from this value. Those small shifts, belonging to a smaller value range, can be stored even in $1$ byte. This conversion led to more than $5\%$ reduction of space requirements.

\subsection{Compression.}
Since there is no need for all landmarks to be concurrently active, we can compress their data blocks. We used the library $zlib$ for this compression, which led to $10\%$ reduction in space.

\subsection{Required space.}
For Berlin, the required space was limited to an average size of less than $14\mathrm{MB}$ per landmark. For storing the time-values of approximate travel-time summaries, we considered $2.64\mathrm{sec}$ as resolution, corresponding to a scale factor $s=1.32$ (when counting time in seconds), which requires $2$ bytes per time-value.
For Germany, the required space was limited to average size of $25.7\mathrm{MB}$ per landmark. For storing the time-values of approximate travel-time summaries, we considered $17.64\mathrm{sec}$ as resolution, corresponding to a scale factor $s=1.32$, and a bucketing factor of $c=15\mathrm{sec}$.

\subsection{Indexing Travel-Time Summaries.}
For retrieving efficiently the required minimum travel-time function $D[\ell,d](t)$ from a landmark $\ell$ to a destination-node $d$, we need also to store an index. Depending on the oracle, we used two types of indices.

\paragraph{Flat Index.}
We maintain a vector of pointers per landmark, one pointer equals per destination. The pointer of destination $v$ provides the address of the $D[\ell,v](t)$ data. The pointers are in ascending order of node ID. The search time is $\Order{1}$ and the required space is $\Order{n|L|}$.

\paragraph{Horn Index.}
For each node $v$ we maintain a vector of pointers. The number of pointers equals to the number of landmarks, from any partition level, which have travel time to the corresponding node $v$ as destination.  Obviously, there exist at least one landmark from the highest partition level. The pointer of the associated landmark $\ell$ provides the address of the $D[\ell,v](t)$ data. The pointers are stored in ascending order of node ID. The search time is $\Order{\log(|L|)}$ and the required space is $\Order{n|L|}$.

%%%%%%%%%%%%%%%%%%%%%%%%%%%%%%%%%%%%%%%%%%%%%%%%%%%%%%%%%%%%%%%%
\section{Live Traffic Reporting}
\label{section:Live-Traffic-Reporting}

% LIVE-TRAFFIC REPORTING MOTIVATION
In a server-side routing service that responds to several queries in real-time, various disruptions may occur ``on the fly'' (e.g., the abrupt and unforeseen congestion, or even blockage of a road segment for half an hour due to a car accident) and have to be taken into account for the affected route plans that have already been suggested or will be suggested in the near future.
% THE PROBLEM...
We thus consider dynamic scenarios where there is a stream of live-traffic reports about abnormal delays on certain road segments (arcs), along with a time-window $[r_s, r_e]$, of typically small duration, in which the disruption occurs.

Our update step involves the recomputation of travel-time summaries for a subset of landmarks in the vicinity of the disruption. In particular, for a disrupted arc $a=uv$ of disruption duration $[r_s,r_e]$, we run a (static) $\alg{Backward\mbox{-}Dijkstra}$ from $u$ under the free-flow metric, with travel time radius of at most $r_e-r_s$. The limited travel time radius is used to trace only the nearest landmarks that may actually be affected by the disruption, leaving unaffected all the ``faraway'' landmarks. The goal is to update as soon as possible the recommendations for the drivers who are close to the area of disruption.
For each affected landmark $\ell$, we consider a \term{disruption-times window} $[t_s, t_e]$, containing the latest departure times from $\ell$ for arriving at the tail $u$ at any time in the interval $[r_s, r_e]$ in which the disruption occurs.
We then compute \emph{temporal} travel-time summaries for each affected landmark and disruption-times window. This computation is conducted as in the preprocessing phase.
Using a $15\mbox{-min}$ radius for the disruptions, we executed $1,000$ live-traffic updates for the instances of Berlin and Germany, each with $2,000$ SR-landmarks. Each disruption has duration at most $2\mathrm{h}$. The average number of affected landmarks was $32$ for Berlin, and only $3$ for Germany (mainly due to sparsity). The corresponding update times for their preprocessed data were $21\mathrm{sec}$ and $37\mathrm{sec}$, respectively.

%%%%%%%%%%%%%%%%%%%%%%%%%%%%%%%%%%%%%%%%%%%%%%%%%%%%%%%%%%%%%%%%
\section{Experimental Evaluation}
\label{section:Experimental-Evaluation}

%%%%%%%%%%%%%%%%%%%%%%%%%%%%%%%%%%%%%%%%%%%%%%%%%%%%%%%%%%%%%%%%
\subsection{Experimental Setup.}
\label{section:Experimental-Setup}

All algorithms were implemented using C++ (gcc version 4.8.2). To support all graph-operations we used the PGL library~\cite{2013-Mali-Michail-Paraskevopoulos-Zaroliagis}. All experiments were executed by a Intel(R) Xeon(R) CPU E5-2643v3 3.40GHz using $128$GB of RAM, on Ubuntu 14.04 LTS.  We used $6$ threads for the parallelism of the preprocessing phase. The query algorithms were executed on a single thread.

%%%%%%%%%%%%%%%%%%%%%%%%%%%%%%%%%%%%%%%%%%%%%%%%%%%%%%%%%%%%%%%%
\subsection{Identity of Instances.}
\label{section:identity-of-instances}

Table~\ref{table:arc-travel-time-statistics} demonstrates the statistics of the arc-travel-time functions in the two instances.

Table~\ref{table:Path-Travel-Time-Statistics} demonstrates some statistics of the minimum-travel-times related to Assumptions~\ref{assumption:Bounded-Travel-Time-Slopes} and \ref{assumption:Bounded-Opposite-Trips}. These statistics are deduced as follows: We run $10,000$ random queries in departure range $R_{dep}$=[9:00-20:00] (densely for rush hours 9:00-12:00). Each query consists of a random pair of nodes $(o, d)$ and a random departure time $t \in R_{dep}$. We compute $D[o,d](t)$ and $D[d,o](t)$ and we set 
$\zeta[o,d] = 	\max\left\{ D[o,d](t) , D[d,o](t) \right\} 
					~/~ \min\left\{D[o,d](t),D[d,o](t)\right\}$. 
$\zeta_{avg} = \sum_{(o,d,t)} \zeta[o,d](t)$ is the average value of $\zeta$ and $\zeta_{\max} = \max_{(o,d,t)} \left\{ \zeta[o,d](t) \right\}$ the maximum value of $\zeta$, over all these random queries. We also present the maximum and minimum observed values of the travel-time slopes $\La_{\max}$ and $-\La_{\min}$ in functions $D[o,d]$, as well as the offset of the computed minimum-travel-time functions.

We now discuss the validity of Assumption~\ref{assumption:growth-of-free-flow-ball-sizes}. Table~\ref{table:Free-Flow-Rank-Blowup} summarizes the observed value of the free-flow blowup that we found in the two instances.

We compute these statistics as folllows: For a set of $5,000$ randomly chosen origins, we grow minimum-travel-time balls by running (static) $\alg{Dijkstra}$ in the \emph{free-flow} metric, until a number $F$ of destinations is settled around each origin. The first column denoted by $FF \ rank$ concerns exactly the different values for $F$ that we consider as the size for these balls. In each of the size-$F$ free-flow balls we compute the maximum min-travel-time $t_{CG}$ of a leaf from its origin, in the \emph{full-congestion} metric this time. We then extend each of the previous balls (again in the \emph{free-flow} metric), until the free-flow radius reaches exactly the value $t_{CG}$. The size of the resulting free-flow ball is denoted by $ExtFF \ Rank$. We provide both the average values $\mathrm{avg}(ExtFF \ Rank)$ and the maximum values $\max(ExtFF \ Rank)$ observed for this parameter, in the instances of Berlin and Germany. As indicated in Table~\ref{table:Free-Flow-Rank-Blowup}, the ratio between $\max(ExtFF \ Rank)$ and $FF \ Rank$ is some constant never exceeding the value $8.3$. The ratio when considering average values never exceeds the value $2$.

%%%%%%%%%%%%%%%%%%%%%%%%%%%%%%%%%%%%%%%%%%%%%%%%%%%%%%%%%%%%%%%%
\subsection{Experimental Evaluation.}

We report here the outcome of our experiments on the instances of Berlin and Germany.
The instance of Berlin consists of $473,253$ nodes and $1,126,468$ arcs. We conducted a preprocessing of the instance that contracts nodes which are not junctions (i.e., of degree $2$ in the undirected graph). We also added some shortcut edges at the endpoints of chains of contracted nodes. This led to an amount of $292,356$ active nodes and $752,362$ active edges in the instance.
As for the instance of Germany, it consists of $4,692,091$ nodes and $11,183,060$ arcs. After the contraction of degree-2 nodes, we got an instance with $3,431,213$ active nodes and $8,554,840$ active edges.

\subsubsection{$\alg{FLAT}$ @ Berlin.}
\label{section:FLAT-AT-BERLIN}

Tables~\ref{table:QUERY-TIME-FLAT-AT-BERLIN} and \ref{table:DIJKSTRA-RANK-FLAT-AT-BERLIN} summarize the performance of the basic query algorithms of $\alg{FLAT}$ with respect to absolute running times and Dijkstra-rank values, respectively, for landmark sets of various sizes. 
% LEGEND FOR TABLES...
The best performance per query algorithm is indicated by highlighted table cells. The last four lines in each table are for the sake of comparison of $\alg{FLAT}$ with $\alg{HORN}$ (see Section~\ref{section:HORN-AT-BERLIN}).
%\textsc{random}
$R$ is for uniform and random landmark selection.
%\textsc{kahip}
$K$ is for selecting the boundary vertices of a \textsc{kahip} partition as landmarks. We have used the version v0.71 of the \textsc{kahip} partitioning software \cite{2014-KAHIP}, exploiting the $\alg{KaFFPa}$ algorithm, with the following parameters: The number of blocks to partition the graph was set to $178$, so that we get (slightly more than) $2,000$ landmarks.
%\textsc{hybrid}
$H$ is for a hybrid partition that initially creates a \textsc{kahip} partition (with half the landmarks) and then randomly chooses additional landmarks within each cell of the partition.
%\textsc{important}-\textsc{random}
$IR$ indicates a variant of $R$ that moves each randomly selected landmark to its closest important node. We have considered as ``important'' those nodes in the Berlin instance which are incident to road segments of category at most $3$.
%\textsc{sparse}-\textsc{random}
$SR$ indicates another variant of randomly selected landmarks, where each newly chosen random landmark excludes its closest $300$ nodes (under the free-flow metric) from being landmarks in the future.
%QUERY ALGORITHMS
As for the query algorithms, we used recursion budget $1$ for $\alg{RQA}$ and we let $\alg{FCA}^+$ settle the $6$ closest landmarks, which is roughly the average number of settled landmarks by $\alg{RQA}$ as well.
\subsubsection{$\alg{HORN}$ @ Berlin.}
\label{section:HORN-AT-BERLIN}

Due to large space requirements, we could handle landmark hierarchies with up to $21,000$ landmarks for the Berlin instance, which seems to be harder than that of Germany\footnote{%
	We observed that the speedups are significantly better in Germany, despite the fact that we consider the same number of landmarks in a larger, by an order of magnitude, network. This is probably due to stronger correlation of time-dependence among different road segments in an urban environment, rather than in a nationwide road network.
}. %FOOTNOTE ENDS 
The average size per landmark in the hierarchy is $2.1$MBytes. For a hierarchy of $10,256$ landmarks the preprocessing of $\alg{HORN}$ took $5.0$ hours, for $20,513$ landmarks it took 9.7 hours, or at most $1.8$sec per landmark in either case. The landmarks in each level of the hierarchy were chosen by the \textsc{random} (HR) and \textsc{sparse}-\textsc{random} (HSR) methods. 
We consider $4$ levels of the hierarchy, according to the sizes of the landmarks' \emph{areas of coverage}, i.e. the number of ``nearby'' destinations for which they possess travel-time summaries. The area of coverage for landmarks of level $4$ is actually the entire graph. These are exactly the ``global'' landmarks which the corresponding variant of $\alg{FLAT}$ would also consider. The landmarks of the other levels have significantly smaller areas of coverage.

In summary, we created four distinct hierarchies, with $10,256$ and $20,513$ landmarks, based on HR and HSR landmark selection methods (cf. Table~\ref{table:HORN-levels}).
The corresponding variants for $\alg{FLAT}$ possess the same ($270$ and $541$, respectively) ``global'' landmarks. 
We created travel-time summaries using only the $\alg{TRAP}$ approximation method, within the subgraph induced by each landmark's area of coverage (extended, as in Assumption~\ref{assumption:growth-of-free-flow-ball-sizes}). 

The experimental results of $\alg{FCA}$ for $R_{270}$, $SR_{270}$, $R_{541}$ and $SR_{541}$ are shown in Tables~\ref{table:QUERY-TIME-FLAT-AT-BERLIN} and \ref{table:DIJKSTRA-RANK-FLAT-AT-BERLIN}. 
Table~\ref{table:HQA-QUERY-TIME+DIJKSTRA-RANK-BERLIN} summarises the experimental evaluation of $\alg{HQA}$. Interestingly,its performance with HR-landmarks is better than that with HSR-landmarks, probably because the ESC criterion seems more effective in the former case.
Table~\ref{table:HQA-vs-FCA-AT-BERLIN} summarizes the comparison of $\alg{HORN}$ with $\alg{FLAT}$. The results for $HR_{10256}$, $HSR_{10256}$, $HR_{20513}$ and $HSR_{20513}$ are compared with the corresponding results of $\alg{FCA}$ for $R_{270}$, $SR_{270}$, $R_{541}$ and $SR_{541}$ (i.e., with the same number of global landmarks), respectively.
There is a significant improvement in query performance (e.g., more than $40\%$ w.r.t. Dijkstra-ranks), but also in quality of the produced solution (by more than $41\%$), at the cost of increasing the space requirements by a factor of $6.44$ at most.

\subsubsection{$\alg{FLAT}$ @ Germany.}
\label{section:FLAT-AT-GERMANY}

We have tried $\alg{FLAT}$ with R-, K-, SR- and SK-landmark sets. Using $6$ threads for the preprocessing phase, the average preprocessing time is less than $90$sec and the average space is up to $25.7$Mbytes, per landmark. We only used $2,000$ landmarks which, as a fraction of the vertex set, is significantly smaller (by an order of magnitude) than in the instance of Berlin. Notably, for the K-landmark set $\alg{FLAT}$'s query performances are now much worse than those for R-, SR- and SK-landmark sets, by almost an order of magnitude. This is probably due to the fact that the number of cells is now too small ($36$) so as to constrain ourselves to roughly $2,000$ landmarks. On the other hand, for the R- SR-, and SK-landmark sets the query performances are quite remarkable, comparable to those for Berlin (by means of speedup), despite the significantly smaller density of the landmarks in the network. Tables~\ref{table:QUERY-TIME-OF-FLAT-IN-GERMANY} and \ref{table:DIJKSTRA-RANK-OF-FLAT-IN-GERMANY} demonstrate the performance of the query algorithms with respect to absolute times and Dijkstra-ranks, respectively. 
E.g., the best speedup against $\alg{TDD}$ is achieved by SK-landmarks, and is  more than $1,531$ in Dijkstra-ranks, and more than $902$ in absolute query-times, with worst-case error at most $1.534\%$.

%%%%%%%%%%%%%%%%%%%%%%%%%%%%%%%%%%%%%%%%%%%%%%%%%%%%%%%%%%%%%%%%%%%%%%%%%%%%%%%%%%%%%%%%%%%%%%
\subsection{Comparison with \cite{2015-Kontogiannis-Michalopoulos-Papastavrou-Paraskevopoulos-Wagner-Zaroliagis}.}

We improve significantly our previous experimental evaluation of $\alg{FLAT}$ in the following sense. We significantly reduce the required preprocessing space, from $70\mathrm{GB}$ down to at most $27\mathrm{GB}$ for the instance of Berlin. This improvement allowed us to also conduct experiments for instance of Germany, which in this new implementation requires space of $51\mathrm{GB}$ for the R-landmarks of size $2,000$. We provide novel landmark sets, which significantly improve the speedup (cf. Table~\ref{table:HQA-vs-FCA-AT-BERLIN}). It is mentioned at this point that, since we have executed our new experiments on a different machine, we re-executed the experiments for $R_{2000}$ and $K_{2000}$. This led to slightly worse running times, compared to the reported values in \cite{2015-Kontogiannis-Michalopoulos-Papastavrou-Paraskevopoulos-Wagner-Zaroliagis}, which are nevertheless more accurate here since we indeed account for the exact execution times, whereas in our previous work we had to exclude the disk-IO accesses due to the limited memory-capabilities that we had at our disposal. 
Moreover, the relative errors and the Dijkstra-rank values in the present paper are better than those reported in  \cite{2015-Kontogiannis-Michalopoulos-Papastavrou-Paraskevopoulos-Wagner-Zaroliagis}, due to several improvements in our code. The best speedup factor is now achieved by SR-landmarks and $\alg{FCA}$: The average query time is $81\mathrm{\mu sec}$, with worst-case relative error $0.771\%$ and speedup more than $1,227$. With respect to the SK-landmarks, $\alg{FCA}$ achieves $83\mathrm{\mu sec}$, with worst-case relative error $0.781\%$ and speedup more than $1,570$.

We also conducted extensive experimental evaluation for random queries of specific ranges, which confirmed our intuition that our query algorithms perform much better for long-range queries rather than for medium-range or short-range queries. This fact indeed motivated our construction of the $\alg{HORN}$ oracle. The evaluation of $\alg{HORN}$ for the instance of Berlin demonstrates that we can achieve remarkable speedups also for short-range and medium-range queries.

Finally, the creation of the temporal travel-time summaries as a response to live-traffic reporting is indeed done very fast, e.g., in less than $21\mathrm{sec}$ for disruptions that would last about $15\mathrm{min}$.

%%%%%%%%%%%%%%%%%%%%%%%%%%%%%%%%%%%%%%%%%%%%%%%%%%%%%%%%%%%%%%%%
\subsection{Comparison with \cite{2013-Batz-Geisberger-Sanders-Vetter}.}

Our main goal in this work is to demonstrate the practicality of $\alg{FLAT}$ and $\alg{HORN}$, which provide \emph{provable} guarantees w.r.t. query times, stretch factors and preprocessing requirements, for large-scale real data sets.
The strong aspects of our oracles are the simplicity of the query algorithms, the remarkably small (optimal in most cases) observed stretches, and the achieved speedups. E.g., for Germany $\alg{FCA}$ responds in $1.269\mathrm{msec}$, achieving a speedup against $\alg{TDD}$ more than $902$ in absolute query-times and $1,531$ w.r.t. the \emph{machine-independent} Dijkstra-ranks, and worst-case error less than $1.534\%$. This performance is clearly very encouraging and quite competitive against the current state-of-art approach of \emph{time-dependent Contraction Hierarchies} \cite{2013-Batz-Geisberger-Sanders-Vetter}.
On the negative side, the preprocessing space and time requirements still remains large, especially when compared to the small space requirements of the TCH-approach of \cite{2013-Batz-Geisberger-Sanders-Vetter}.

%%%%%%%%%%%%%%%%%%%%%%%%%%%%%%%%%%%%%%%%%%%%%%%%%%%%%%%%%%%%%%%%
\section{Conclusions and Future Work}
\label{section:Conclusions}

We provided an extensive experimental evaluation of landmark-based oracles for time-dependent road networks. 
We are currently exploring different landmark sets that will achieve even better speedups and/or approximation guarantees. As a first step, inspired by the \emph{sparsification} in the R-landmark set, we also explored techniques for sparsifying also the \textsc{kahip} landmark sets. Preliminary results for Berlin have demonstrated remarkable results. E.g., we achieved average response times of $83\mu\mathrm{sec}$, worst-case error of $0.781\%$ and Dijkstra-rank speedup more than $1,570$ for the Berlin instance.

In case that space is a main concern, we observed the full scalability in the treadoffs of our oracles between space and query-responses. E.g., consuming space $3.2\mathrm{GB}$ we can achieve query-response times $0.73\mathrm{msec}$, relative error $2.198\%$, for the Berlin instance. We are currently experimenting with more sophisticated landmark schemes to further improve the space vs. query-responses tradeoff.

We are finally exploring further improvements in the compression schemes, and the exploitation of parallelism, not only as a simple load-balancing scheme, but also algorithmically, which will further reduce the requirements for preprocessing space and time, and live-traffic updating.

%%%%%%%%%%%%%%%%%%%%%%%%%%%%%%%%%%%%%%%%%%%%%%%%%%%%%%%%%%%%%%%%
\bibliographystyle{plain}

%\bibliography{../../../Shortest-Paths}

%%%%%%%%%%%%%%%%%%%%%%%%%%%%%%%%%%%%%%%%%%%%%%%%%%%%%%%%%%%%%%%%

\clearpage

\section*{List of Tables}

%%%%%%%%%%%%%%%%%%%%%%%%%%%%%%%%%%%%%%%%%%%%%%%%%%%%%%%%%%%%%%%%
%%                            appendix.tex                    %%
%%%%%%%%%%%%%%%%%%%%%%%%%%%%%%%%%%%%%%%%%%%%%%%%%%%%%%%%%%%%%%%%

%%%%%%%%%%%%%%%%%%%%%%%%%%%%%%%%%%%%%%%%%%%%%%%%%%%%%%%%%%%%%%
%\section{Summary of Notation}
%\label{section:table-of-notation}

\begin{table}[htb]
\center
\begin{footnotesize}
\begin{tabular}{|>{\columncolor{table subheader}}c|p{13cm}|}
\hline
\rowcolor{table header}
Symbol & Description
\\ \hline
$[k]$ & The set of integers $\{1,2,\ldots,k\}$.
\\ \hline
$G=(V,A)$ & The graph representing the underlying road network.
	%\newline
	$n = |V|$ and $m=|A|$.
\\ \hline
$diam(G,D)$ & The diameter of $G$ under an arc-cost metric $D$.
\\ \hline
$\mathcal{P}_{o,d}$ & Set of $od$-paths in $G$.
\\ \hline
$p \bullet q$ & The concatenation of the $ux$-path $p$ with the $xv$-path $q$ at vertex $x$.
\\ \hline
$ASP[o,d](t_o)$ & Set of $(1+\eps)$-approximations of minimum-travel-time $od$-paths in $G$, for given departure-time $t_o\geq 0$.
\\ \hline
$SP[o,d](t_o)$ & Set of minimum-travel-time $od$-paths in $G$, for given departure-time $t_o\geq 0$.
\\ \hline
$B[v](t_v)$ & A ball growing from $(v,t_v)\in V\times[0,T)$, in the time-dependent metric, until either the destination $d$ is reached or the closest landmark $\ell_v$ from $(v,t_v)$ is settled.
\\ \hline
$B[v;F](t_v)$ & A ball growing from $(v,t_v)\in V\times[0,T)$, in the time-dependent metric, of size $F\in\naturals$.
\\ \hline
$\overline{B}[v;F]$ / $\underline{B}[v;F]$ & A ball growing from $v\in V$, in the full-congestion / free flow metric, of (integer) size $F\in\naturals$.
\\ \hline
$\overline{B}[v;R]$ / $\underline{B}[v;R]$ & A ball growing from $v\in V$, in the full-congestion / free flow metric, of (scalar) radius $R>0$.
\\ \hline
$B'[v;F](t_v)$ & A ball growing from $(v,t_v)\in V\times[0,T)$, in the time-dependent metric, of size $F\polylog(F)$, according to Assumption~\ref{assumption:growth-of-free-flow-ball-sizes}.
\\ \hline
$d[a](t)$ & The limited-window arc-travel-time function for arc $a\in A$, with departure-time $t\in [0,T)$ for some \emph{constant} time-period $T>0$ (e.g., a single day).
\\ \hline
$M_a$ & Maximum possible travel-time ever seen at arc $a$.
\\ \hline
$M$ & Maximum arc-travel-time ever seen in any arc.
\\ \hline
$D[a](t)$ & Periodic arc-travel-time function for arc $a\in A$, with domain $t\in [0,\infty)$.
\\ \hline
$Arr[a](t)$ & The arc-arrival-time function for arc $a\in A$.
\\ \hline
$D[o,d]$ & Minimum-travel-time \emph{function}, from $o$ to $d$.
\\ \hline
$\Gamma[o,d]$ & Dijkstra-ranks \emph{function}, from $o$ to $d$.
\\ \hline
$D_{\max}[o,d]$ / $D_{\min}[o,d]$ & The maximum and minimum value of $D[o,d]$.
\\ \hline
$\overline{D}[a]$ / $\underline{D}[a]$ & Travel-times of $a$ in \emph{full-congestion} and \emph{free-flow} metrics, respectively.
\\ \hline
$\overline{\De}[o,d]$ / $\underline{\De}[o,d]$ & An upper-approximating / lower-approximating function to $D[o,d]$.
\\ \hline
$Arr[o,d]$ & Earliest-arrival-time \emph{function}, from $o$ to $d$.
\\ \hline
$t_u ~~ (t_v)$ & Departure-time from the tail $u$ (arrival-time at the head $v$) for the arc $uv\in A$.
\\ \hline
$TDSP(o,d,t_o)$ & The problem of finding an min-cost $od$-path, given a departure-time $t_o$.
\\ \hline
$TDSP(o,\star,t_o)$ & The problem of finding a min-cost paths tree from $o$, given a departure-time $t_o$.
\\ \hline
$TDSP(o,d)$ & The problem of constructing a succinct representation of min-cost $od$-paths \emph{function}.
\\ \hline
$K_a$ & Number of breakpoints in the arc-travel-time function $D[a]$.
\\ \hline
$K$ & Total number of breakpoints in the arc-travel-time functions.
\\ \hline
$K_{\max}$ & The maximum number of breakpoints, among the arc-travel-time functions.
\\ \hline
$K^*$ & Total number of \emph{concavity-spoiling} breakpoints in the arc-travel-time functions.
\\ \hline
$\La_{\max}$ & Maximum slope among minimum-travel-time functions.
\\ \hline
$\La_{\min}$ & Absolute value of minimum slope among minimum-travel-time functions.
\\ \hline
$\zeta$ & ratio of min-travel-times in opposite directions between two vertices for any specific departure-time.
\\ \hline
$r$ & The recursion budget for $\alg{RQA}$ and $\alg{HQA}$.
\\ \hline
%
%%% $F_{\max}$ & Maximum number of legs in a minimum-travel-time function from the origin $o$ to any vertex in the network. \\ \hline
%%% $F_{o,d}$ & Number of legs in the pwl function $D[o,d]$. \\ \hline
%%% $F^*$ & Total number of legs in all minimum-travel-time functions. \\ \hline
%
$\alg{BIS}$ & The bisection approximation method for minimum-travel-time functions.
\\ \hline
$\alg{TRAP}$ & The trapezoidal approximation method for minimum-travel-time  functions.
\\ \hline
$\alg{FCA}$ & The \emph{Forward Constant Approximation} query algorithm.
\\ \hline
$\alg{RQA}$ & The \emph{Recursive Query Algorithm}.
\\ \hline
$\alg{HQA}$ & The \emph{Hierarchical Query Algorithm}, based on a hierarchy of landmarks.
\\ \hline
$\alg{FLAT}$ & The oracle that uses landmarks possessing summaries towards all possible destinations.
\\ \hline
$\alg{HORN}$ & The oracle that uses a hierarchy of landmarks and the $\alg{HQA}$ query algorithm.
\\ \hline
\end{tabular}
\end{footnotesize}
\caption{\label{table:notation} Summary of notation.}
\end{table}

\clearpage

%%%%%%%%%%%%%%%%%%%%%%%%%%%%%%%%%%%%%%%%%%%%%%%%%%%%%%%%%%%%%%%%
%\section{Identity of Instance and Verification of Assumptions}

\begin{table}[htb]
\center
\begin{tabular}{|>{\columncolor{table subheader}}r|c|c|}
\hline
\rowcolor{table header}
${}_{\mbox{PARAMETER}} ~\setminus~ {}^{\mbox{INSTANCE}}$
													& Berlin (TomTom) & Germany (PTV AG) \\ \hline
\#Nodes											& 473,253				& 4,692,091       \\ \hline
\#Edges											& 1,126,468			& 11,183,060      \\ \hline
Time Period										& {$24$h (Tue)} 	& 24h {(Tue-Wed-Thu)}
\\ \hline
$\la_{\max}$										& 0.017          & 0.130           \\ \hline
$-\la_{\min}$										& -0.013         & -0.130          \\ \hline
\#Arcs with constant traversal-times		& 924,254        & 10,310,234      \\ \hline
\#Arcs with non-constant traversal-times	& 20,2214        & 872,826         \\ \hline
Min \#Breakpoints  								& 4              & 5               \\ \hline
Avg \#Breakpoints								& 10.4           & 16.3            \\ \hline
Max \#Breakpoints								& 125            & 52              \\ \hline
Total \#Breakpoints                       & 3,234,213      & 25,424,506      \\ \hline
\end{tabular}
\caption{\label{table:arc-travel-time-statistics}%
	Arc-travel-time function statistics. $\la_{\max}$ and $-\la_{\min}$ are the maximium and minimum slope among the arc-traversal-time functions.
}%CAPTION
\end{table}

\begin{table}[htb!]
\center
\begin{tabular}{|>{\columncolor{table subheader}}r|r|r|}
\hline
\rowcolor{table header}
${}_{\mbox{PARAMETER}} ~\setminus~ {}^{\mbox{INSTANCE}}$
										& Berlin (TomTom)	& Germany (PTV AG) 
\\ \hline
$\zeta_{avg}$						& 1.008		& 1.004   
\\ \hline
$\zeta_{\max}$						& 1.189		& 1.050   
\\ \hline
$\La_{\max}$    					& 0.190		& 0.216   
\\ \hline
$-\La_{\min}$     					& -0.150	& -0.197  
\\ \hline
Max Offset    						& 46,039	& 51,850 
\\ \hline
Min Offset    						& -1,472	& -6,331 
\\ \hline
Max Min-Travel-Time					& 43,797	& 48,382 
\\ \hline
\end{tabular}
\caption{\label{table:Path-Travel-Time-Statistics}%
	Min-cost path-travel-time statistics.
}%CAPTION
\end{table}

\begin{table}[htb!]
\center
\begin{tabular}{|>{\columncolor{table subheader}}r|r|r|r|r|r|r|}
\hline
\cellcolor{table header}
& \multicolumn{3}{c|}{\cellcolor{table header} $ExtFF \ Rank$ @ Berlin}		
& \multicolumn{3}{c|}{\cellcolor{table header} $ExtFF \ Rank$ @ Germany}
\\ \hline
\rowcolor{table subheader}
$FF$ $Rank$
& avg 
& max  	
& blow-up $(\%)$
& avg
& max  	
& blow-up $(\%)$
\\ \hline
100     	& 136.9        	& 671   		& $6.710$
			& 135.9    		& 670			& $6.700$
\\ \hline
1,000    	& 1,532.7      	& 3,898		& $3.898$
			& 1,661.7			& 8,299 		& \cellcolor{bold cell color} $8.299$
\\ \hline
4,000    	& 6,315.0      	& 15,552		& $3.888$
			& 7,110.2   		& 16,024		& $4.006$
\\ \hline
6,000    	& 9,669.1      	& 25,951		& $4.326$
			& 10,783.3  		& 29,145 		& $4.858$
\\ \hline
10,000   	& 16,142.8     	& 36,744		& $3.675$
			& 17,890.1  		& 42,415		& $4.242$
\\ \hline
\end{tabular}
\caption{\label{table:Free-Flow-Rank-Blowup}
	Free-flow blow-up statistics for the instances of Berlin and Germany. The worst-case blow-up factor is less than $8.3$ times.}
\end{table}

\clearpage 

%%%%%%%%%%%%%%%%%%%%%%%%%%%%%%%%%%%%%%%%%%%%%%%%%%%%%%%%%%%%%%%%
%\section{Detailed Experimental Results}

\begin{table*}[htb!]
\centering
\begin{footnotesize}
\begin{tabular}{r|c|c|c|c|c|c|c|c|}
\hhline{*{1}{~|}*{8}{-|}}
	& \multicolumn{2}{c|}{\cellcolor{table header} $\alg{TDD}$}
	& \multicolumn{2}{c|}{\cellcolor{table header} $\alg{FCA}$}	
	& \multicolumn{2}{c|}{\cellcolor{table header} $\alg{FCA}^+(6)$}	
	& \multicolumn{2}{c|}{\cellcolor{table header} $\alg{RQA}$}
\\
\hhline{*{1}{~|}*{8}{-|}}
	& \cellcolor{table subheader} \shortstack{\\Time \\ (msec)}
	& \cellcolor{table subheader} \shortstack{\\Rel.Error \\ \%}
	& \cellcolor{table subheader} \shortstack{\\Time \\ (msec)}
	& \cellcolor{table subheader} \shortstack{\\Rel.Error \\ \%}
	& \cellcolor{table subheader} \shortstack{\\Time \\ (msec)}
	& \cellcolor{table subheader} \shortstack{\\Rel.Error \\ \%}
	& \cellcolor{table subheader} \shortstack{\\Time \\ (msec)}
	& \cellcolor{table subheader} \shortstack{\\Rel.Error \\ \%}
\\
\hhline{*{9}{-|}}
\hhline{*{9}{-|}}
$R_{2000}$ & \multirow{2}{*}{$92.906$} & \multirow{2}{*}{$0$}
& $0.100$ & $0.969$ & $0.527$ & $0.405$ & $0.519$ & $0.679$
\\
%\cline{4-9}
\hhline{*{3}{~|}*{6}{-|}}
$K_{2000}$ & &
& $0.115$ & $1.089$ 
& \cellcolor{bold cell color}$0.321$ & $0.405$ 
& \cellcolor{bold cell color}$0.376$ & $0.523$
\\
%\cline{4-9}
\hhline{*{3}{~|}*{6}{-|}}
$H_{2000}$ & &
& $0.102$ & $0.886$ & $0.523$ & $0.332$ & $0.445$ & $0.602$
\\
%\cline{4-9}
\hhline{*{3}{~|}*{6}{-|}}
$IR_{2000}$ & &
& $0.086$ & $0.923$ 
& $0.489$ & $0.379$ 
& $0.473$ & $0.604$
\\
%\cline{4-9}
\hhline{*{3}{~|}*{6}{-|}}
$SR_{2000}$ & &
&\cellcolor{alert cell color}$0.081$
&\cellcolor{alert cell color}$0.771$ 
& $0.586$ & $0.317$ 
& $0.443$ & $0.611$
\\
\hhline{*{3}{~|}*{6}{-|}}
  $SK_{2000}$ & &
& $0.083$
& $0.781$ 
& $0.616$ & \cellcolor{bold cell color}$0.227$ 
& $0.397$ & \cellcolor{bold cell color}$0.464$
\\
\hhline{*{9}{-|}}
\hhline{*{9}{-|}}
$R_{541}$ 	& &
&\cellcolor{bold cell color}$0.326$ 
& $1.854$ 
&\cellcolor{bold cell color}$1.887$ 
& $0.693$ 
&\cellcolor{bold cell color}$1.904$ 
& $1.610$
\\
%\cline{4-9}
\hhline{*{3}{~|}*{6}{-|}}
$SR_{541}$ & &
& $0.451$ 
&\cellcolor{bold cell color}$1.638$ 
& $3.252$ 
&\cellcolor{bold cell color}$0.614$ 
& $2.856$ 
&\cellcolor{bold cell color}$1.531$
\\
\hhline{*{9}{-|}}
\hhline{*{9}{-|}}
$R_{270}$ 	& &
&\cellcolor{bold cell color}$0.639$ 
& $2.583$ 
&\cellcolor{bold cell color}$3.707$ 
& $0.881$ 
&\cellcolor{bold cell color}$3.842$ 
& $2.482$
\\
%\cline{4-9}
\hhline{*{3}{~|}*{6}{-|}}
$SR_{270}$ & &
&$0.730$ 
&\cellcolor{bold cell color}$2.198$ 
& $4.491$ 
&\cellcolor{bold cell color}$0.745$ 
& $4.271$ 
&\cellcolor{bold cell color}$2.336$
\\
\hhline{*{1}{~|}*{8}{-|}}
\end{tabular}
\caption{\label{table:QUERY-TIME-FLAT-AT-BERLIN}
	Performance of $\alg{FCA}$, $\alg{FCA}^+(6)$ and $\alg{RQA}$, w.r.t. the \emph{running times} and \emph{relative errors}, at $2.64$sec resolution, for a query set of $10,000$ \emph{random queries} in Berlin. 
%The last four rows concern $\alg{FLAT}$'s performance w.r.t. the ``global'' landmarks of $\alg{HORN}$.
}%CAPTION
\end{footnotesize}
\end{table*}

\begin{table}[htb!]
\centering
\begin{footnotesize}
\begin{tabular}{r|c|c|r|r|r|r|r|r|}
\hhline{*{1}{~|}*{8}{-|}}
	& \multicolumn{2}{c|}{\cellcolor{table header} $\alg{TDD}$}
	& \multicolumn{2}{c|}{\cellcolor{table header} $\alg{FCA}$}	
	& \multicolumn{2}{c|}{\cellcolor{table header} $\alg{FCA}^+(6)$}	
	& \multicolumn{2}{c|}{\cellcolor{table header} $\alg{RQA}$}
\\
\hhline{*{1}{~|}*{8}{-|}}
	& \cellcolor{table subheader} Rank
	& \cellcolor{table subheader} Speedup
	& \cellcolor{table subheader} Rank
	& \cellcolor{table subheader} Speedup
	& \cellcolor{table subheader} Rank
	& \cellcolor{table subheader} Speedup
	& \cellcolor{table subheader} Rank
	& \cellcolor{table subheader} Speedup
\\
\hhline{*{9}{-|}}
\hhline{*{9}{-|}}
$R_{2000}$ & \multirow{2}{*}{$146,022$} & \multirow{2}{*}{$1$}
& $150$ & $973.480$ & $877$ & $166.502$ & $925$ & $157.862$
\\
%\cline{4-9}
\hhline{*{3}{~|}*{6}{-|}}
$K_{2000}$ & &
& $190$ 
& $768.537$ 
& $866$ 
& $168.616$ 
& $670$ 
& $217.943$
\\
%\cline{4-9}
\hhline{*{3}{~|}*{6}{-|}}
$H_{2000}$ & &
& $154$ 
& $948.195$ 
& $851$ 
& $171.589$ 
& $777$ 
& $187.931$
\\
%\cline{4-9}
\hhline{*{3}{~|}*{6}{-|}}
$IR_{2000}$ & &
& $135$ & $1,081.644$ & $823$ 
& $177.426$ 
& $839$ 
& $174.043$
\\
%\cline{4-9}
\hhline{*{3}{~|}*{6}{-|}}
$SR_{2000}$ & &
& $119$ 
& $1,227.075$ 
& $952$ 
& $153.384$ 
& $776$ 
& $188.173$
\\
\hhline{*{3}{~|}*{6}{-|}}
$SK_{2000}$ & &
& $93$ 
& \cellcolor{alert cell color} $1,570.129$ 
& $755$ 
& \cellcolor{bold cell color} $193.406$ 
& $501$ 
& \cellcolor{bold cell color} $291.461$
\\
\hhline{*{9}{-|}}
\hhline{*{9}{-|}}
$R_{541}$ 	& &
& $545$ 
&\cellcolor{bold cell color}$267.930$ 
& $3,178$ 
&\cellcolor{bold cell color}$45.947$ 
& $3,406$ 
&\cellcolor{bold cell color}$42.872$
\\
%\cline{4-9}
\hhline{*{3}{~|}*{6}{-|}}
$SR_{541}$	& &
& $638$ 
& $228.874$ 
& $3,684$ 
& $39.637$ 
& $3,950$ 
& $36.967$
\\
\hhline{*{9}{-|}}
\hhline{*{9}{-|}}
$R_{270}$ 	& &
& $1,075$ 
&\cellcolor{bold cell color}$135.834$ 
& $6,198$ 
&\cellcolor{bold cell color}$23.559$ 
& $6,702$ 
&\cellcolor{bold cell color}$21.788$
\\
%\cline{4-9}
\hhline{*{3}{~|}*{6}{-|}}
$SR_{270}$	& &
& $1,195$ 
& $122.194$ 
& $7,362$ 
& $19.835$ 
& $7,398$ 
& $19.738$
\\
\hhline{*{1}{~|}*{8}{-|}}
\end{tabular}
\caption{\label{table:DIJKSTRA-RANK-FLAT-AT-BERLIN}
	Performance of $\alg{FCA}$, $\alg{FCA}^+(6)$ and $\alg{RQA}$, w.r.t. \emph{Dijkstra ranks}, at $2.64$sec resolution, for a query set of $10,000$ \emph{random queries} in Berlin. 
%The last four rows concern $\alg{FLAT}$'s performance w.r.t. the ``global'' landmarks of $\alg{HORN}$.
}%CAPTION
\end{footnotesize}
\end{table}

\begin{table}[htb]
\centering
\begin{footnotesize}
\begin{tabular}{|r|r|r|r|r|r|}
\hline
\cellcolor{table header} Level & \multicolumn{2}{c|}{\cellcolor{table header} Size of Levels} 	& \cellcolor{table header} Area of coverage & \multicolumn{2}{c|}{\cellcolor{table header} Excluded Ball Size (for $HSR$)}
\\ 
\cline{2-3}\cline{5-6}
\cellcolor{table header} 
& \cellcolor{table subheader} $|L| = 10,256$	
& \cellcolor{table subheader} $|L| = 20,513$	
& \cellcolor{table header} 
& \cellcolor{table subheader} $|L| = 10,256$	
& \cellcolor{table subheader} $|L| = 20,513$
\\ \hline
$L_1$		& $7,685$				& $15,370$			& $1,274$				& $35$					& $15$
\\ \cline{2-6}
$L_2$		& $1,604$				& $3,208$ 			& $29,243$			& $150$				& $80$
\\ \cline{2-6}
$L_3$		& $697$				& $1,394$				& $154,847$			& $350$				& $180$
\\ \cline{2-6}
$L_4$		& $270$				& $541$				& $292,356$ 			& $800$				& $400$
\\ \hline
\end{tabular}
\caption{\label{table:HORN-levels}
	Landmark hierarchies for $\alg{HORN}$, based on HR and HSR landmark selection methods, for the Berlin instance.
}%CAPTION
\end{footnotesize}
\end{table}
\begin{table}[htb!]
\centering
\begin{footnotesize}
\begin{tabular}{r|c|c|c|c|r|r|r|r|}
\hhline{*{1}{~|}*{8}{-|}}
	& \multicolumn{4}{c|}{\cellcolor{table header} $\alg{TDD}$}
	& \multicolumn{4}{c|}{\cellcolor{table header} $\alg{HQA}$}
\\
\hhline{*{1}{~|}*{8}{-|}}
	& \cellcolor{table subheader} \shortstack{\\Time \\ (msec)}
	& \cellcolor{table subheader} \shortstack{\\Rel.Error \\ \%}
	& \cellcolor{table subheader} \shortstack{\\Rank \\[8pt]}
	& \cellcolor{table subheader} \shortstack{\\Speedup \\[6pt]}
	& \cellcolor{table subheader} \shortstack{\\Time \\ (msec)}
	& \cellcolor{table subheader} \shortstack{\\Rel.Error \\ \%}
	& \cellcolor{table subheader} \shortstack{\\Rank \\[8pt]}
	& \cellcolor{table subheader} \shortstack{\\Speedup \\[6pt]}
\\
\hhline{*{9}{-|}}
\hhline{*{9}{-|}}
$HR_{10256}$
& \multirow{2}{*}{$92.906$} & \multirow{2}{*}{$0$}
& \multirow{2}{*}{$146,022$} & \multirow{2}{*}{$1$}
&\cellcolor{bold cell color}$0.354$ 
& $1.499$
&\cellcolor{bold cell color}$636$ 
&\cellcolor{bold cell color}$229.594$
\\
%\cline{6-9}
\hhline{*{5}{~|}*{4}{-|}}
$HSR_{10256}$ & & & &
& $0.436$ 
&\cellcolor{bold cell color}$1.409$
& $721$ & $202.527$
\\
\hhline{*{9}{-|}}
\hhline{*{9}{-|}}
$HR_{20513}$ & & & &
&\cellcolor{bold cell color}$0.217$ 
& $1.051$
&\cellcolor{bold cell color}$324$ 
&\cellcolor{bold cell color}$450.685$
\\
%\cline{6-9}
\hhline{*{5}{~|}*{4}{-|}}
$HSR_{20513}$ & & & &
& $0.314$ 
&\cellcolor{bold cell color}$0.919$
& $378$ 
& $386.302$
\\
\hhline{*{1}{~|}*{8}{-|}}
\end{tabular}
\caption{\label{table:HQA-QUERY-TIME+DIJKSTRA-RANK-BERLIN}
	Performance of $\alg{HQA}$, w.r.t. the \emph{running times}, \emph{relative errors} and \emph{Dijkstra ranks}, at $2.64$sec resolution, for a query set of $10,000$ \emph{random queries} in Berlin.}
\end{footnotesize}
\end{table}

\begin{table}[htb!]
\centering
\begin{footnotesize}
\begin{tabular}{r|c|c|c|c|}
\hhline{*{1}{~}*{4}{-|}}
\rowcolor{table header}
\cellcolor{white}
& \multicolumn{3}{c|}{\cellcolor{table header} Improvement in}
& Deterioration in
\\
\hhline{*{1}{~}*{4}{-|}}
\rowcolor{table subheader}
\cellcolor{white}
& Query Times	($\%$) 
& Worst-case Relative Error ($\%$) 
& Dijkstra Ranks ($\%$)  
& Space (times)
\\
\hhline{*{5}{-|}}
\hhline{*{5}{-|}}
%\cline{2-5}
$R_{270}$ vs $HR_{10256}$	
& $44.60$	& $41.96$	& $40.83$	& $6.089$
\\
\hhline{*{1}{~}*{4}{-|}}
$SR_{270}$ vs $HSR_{10256}$
& $40.27$ 
& $35.89$	
& $39.66$	
& $6.407$
\\
\hhline{*{5}{-|}}
\hhline{*{5}{-|}}
$R_{541}$ vs $HR_{20513}$	
& $33.43$	
& $43.31$	
& $40.55$	& 
$6.195$
\\
\hhline{*{1}{~}*{4}{-|}}
$SR_{541}$ vs $HSR_{20513}$	
& $30.37$	
& $43.89$	
& $40.75$	
& $6.438$
\\
\hhline{*{1}{~}*{4}{-|}}
\end{tabular}
\caption{\label{table:HQA-vs-FCA-AT-BERLIN}
	Comparison of $\alg{HQA}$ versus $\alg{FCA}$ in Berlin. Two global-landmark sizes ($270$ and $541$) and two landmark-selection methods (\textsc{random} and \textsc{sparse-random}) are considered.}
\end{footnotesize}
\end{table}

\begin{table}[htb]
\centering
\begin{footnotesize}
\begin{tabular}{r|c|c|r|r|r|r|r|r|}
\hhline{*{1}{~|}*{8}{-|}}
	& \multicolumn{2}{c|}{\cellcolor{table header} $\alg{TDD}$}
	& \multicolumn{2}{c|}{\cellcolor{table header} $\alg{FCA}$}	
	& \multicolumn{2}{c|}{\cellcolor{table header} $\alg{FCA}^+(6)$}	
	& \multicolumn{2}{c|}{\cellcolor{table header} $\alg{RQA}$}
\\
\hhline{*{1}{~|}*{8}{-|}}
	& \cellcolor{table subheader} \shortstack{\\Time \\ (msec)}
	& \cellcolor{table subheader} \shortstack{\\Rel.Error \\ \%}
	& \cellcolor{table subheader} \shortstack{\\Time \\ (msec)}
	& \cellcolor{table subheader} \shortstack{\\Rel.Error \\ \%}
	& \cellcolor{table subheader} \shortstack{\\Time \\ (msec)}
	& \cellcolor{table subheader} \shortstack{\\Rel.Error \\ \%}
	& \cellcolor{table subheader} \shortstack{\\Time \\ (msec)}
	& \cellcolor{table subheader} \shortstack{\\Rel.Error \\ \%}
\\
\hhline{*{9}{-|}}
\hhline{*{9}{-|}}
$R_{2000}$ 
& \multirow{2}{*}{$1,145.060$} 
& \multirow{2}{*}{$0$}
& $1.532$ 
& $1.567$ 
& \cellcolor{bold cell color} $8.529$ 
& $0.742$ 
& $9.219$ 
& $1.502$
\\
%\cline{4-9}
\hhline{*{3}{~|}*{6}{-|}}
$K_{2000}$ & &
& $10.455$ & $2.515$ & $15.209$ & $1.708$ & $30.577$ & $2.343$
\\
%\cline{4-9}
\hhline{*{3}{~|}*{6}{-|}}
$SR_{2000}$ & &
& $1.275$ 
& \cellcolor{alert cell color}$1.444$ 
& $9.952$ 
& \cellcolor{bold cell color}$0.662$ 
& $9.011$ 
& \cellcolor{bold cell color}$1.412$
\\
\hhline{*{3}{~|}*{6}{-|}}
$SK_{2000}$ & &
& \cellcolor{alert cell color} $1.269$ 
& $1.534$ 
& $9.689$ 
& $0.676$ 
& \cellcolor{bold cell color}$7.653$ 
& $1.475$
\\
\hhline{*{1}{~|}*{8}{-|}}
\end{tabular}
\caption{\label{table:QUERY-TIME-OF-FLAT-IN-GERMANY}
	Performance of $\alg{FCA}$, $\alg{FCA}^+(6)$ and $\alg{RQA}$, w.r.t. the \emph{running times} and \emph{relative errors}, at $17.64$sec resolution, for a query set of 10,000 \emph{random queries} in Germany.}
\end{footnotesize}
\end{table}

\begin{table}[htb]
\centering
\begin{footnotesize}
\begin{tabular}{r|c|c|r|r|r|r|r|r|}
\hhline{*{1}{~|}*{8}{-|}}
	& \multicolumn{2}{c|}{\cellcolor{table header} $\alg{TDD}$}
	& \multicolumn{2}{c|}{\cellcolor{table header} $\alg{FCA}$}	
	& \multicolumn{2}{c|}{\cellcolor{table header} $\alg{FCA}^+(6)$}	
	& \multicolumn{2}{c|}{\cellcolor{table header} $\alg{RQA}$}
\\
\hhline{*{1}{~|}*{8}{-|}}
	& \cellcolor{table subheader} Rank
	& \cellcolor{table subheader} Speedup
	& \cellcolor{table subheader} Rank
	& \cellcolor{table subheader} Speedup
	& \cellcolor{table subheader} Rank
	& \cellcolor{table subheader} Speedup
	& \cellcolor{table subheader} Rank
	& \cellcolor{table subheader} Speedup
\\
\hhline{*{9}{-|}}
\hhline{*{9}{-|}}
$R_{2000}$ & \multirow{2}{*}{$1,717,793$} & \multirow{2}{*}{$1$}
& $1,659$ & $1,035.439$ & $10,159$ & $169.091$ & $11,045$ & $155.527$
\\
%\cline{4-9}
\hhline{*{3}{~|}*{6}{-|}}
$K_{2000}$ & &
& $9,302$ & $184.669$ & $15,373$ & $111.741$ & $30,137$ & $56.999$
\\
%\cline{4-9}
\hhline{*{3}{~|}*{6}{-|}}
$SR_{2000}$ & &
& $1,277$ 
& $1,345.178$ 
& $9,943$ 
& $172.764$ 
& $9,182$ 
& $187.082$
\\
\hhline{*{3}{~|}*{6}{-|}}
$SK_{2000}$ & &
& \cellcolor{alert cell color}$1,122$ 
& \cellcolor{alert cell color}$1,531.010$ 
& \cellcolor{bold cell color}$9,000$ 
& \cellcolor{bold cell color}$190.866$ 
& \cellcolor{bold cell color}$7,975$ 
& \cellcolor{bold cell color}$215.397$
\\
\hhline{*{1}{~|}*{8}{-|}}
\end{tabular}
\caption{\label{table:DIJKSTRA-RANK-OF-FLAT-IN-GERMANY}
	Query performance of $\alg{FCA}$, $\alg{FCA}^+(6)$ and $\alg{RQA}$, w.r.t. Dijkstra ranks, at $17.64$sec resolution, for a query set of $10,000$ \emph{random queries} in Germany.}
\end{footnotesize}
\end{table}

\end{document}